\documentclass[11pt,a4paper]{article}
\pdfoutput=1
\usepackage{jheppub}

\usepackage{color}
\usepackage{amsmath}
\usepackage{pifont}   
\usepackage{verbatim}       
\usepackage{acronym} 

\usepackage{amsfonts}
\usepackage{amssymb}
\usepackage{mathrsfs} 
\usepackage{graphicx}
\usepackage{multirow} 
\usepackage{slashed} 
\usepackage{array}

\usepackage{graphicx}
\usepackage{epsfig}
\usepackage{lscape}
\usepackage{rotating}
\usepackage{epsf}
\usepackage{bm}
\usepackage{booktabs}
\usepackage{array}
\usepackage{caption}
\usepackage{subcaption}
\usepackage[utf8]{inputenc}
\usepackage{float}

\usepackage{multirow}
\usepackage{array,booktabs,colortbl,multirow}
\usepackage{colortbl,xcolor,colordvi,color}
\usepackage{rotating}
\allowdisplaybreaks

\newcolumntype{L}{>{$}l<{$}} 
\newcommand{\ba}{\begin{eqnarray}}
\newcommand{\ea}{\end{eqnarray}}

\newcommand{\be}{\begin{equation}}
\newcommand{\ee}{\end{equation}}
\newcommand{\eq}[1]{Eq.~(\ref{#1})}

\newcommand*{\email}[1]{\href{mailto:#1}{\nolinkurl{#1}} } 
\usepackage{hyperref}
\hypersetup{
colorlinks = true,
linkcolor = black,
citecolor={blue}
}
\newcommand{\cevns}{{CE$\nu$NS~}}

\title{Probing light vector mediators with coherent scattering at future facilities}

\author[a]{E.~Bertuzzo,}
\author[b]{G.~Grilli~di~Cortona,}
\author[a]{and L.~Magno~Dantas~Ramos}

\affiliation[a]{Instituto de F\'{i}sica, Universidade de S\~{a}o Paulo, C.P. 66.318, 05315-970 S\~{a}o Paulo, Brazil}
\affiliation[b]{Istituto Nazionale di Fisica Nucleare, Laboratori Nazionali di Frascati, C.P. 13, 00044 Frascati, Italy}

\emailAdd{bertuzzo@if.usp.br}
\emailAdd{grillidc@lnf.infn.it}
\emailAdd{lucas.magno.ramos@usp.br}

\abstract{Future experiments dedicated to the detection of Coherent Elastic Neutrino-Nucleus Scattering may be powerful tools in probing light new physics. In this paper we study the sensitivity on light $Z'$ mediators of two proposed experiments: a directional low pressure Time Projection Chamber detector, $\nu$BDX-DRIFT, that will utilize neutrinos produced at the Long Baseline Neutrino Facility, and several possible experiments to be installed at the European Spallation Source. We compare the results obtained with existing limits from fixed-target, accelerator, solar neutrino and reactor experiments. Furthermore, we show that these experiments have the potential to test unexplored regions that, in some case, could explain the anomalous magnetic moment of the muon or peculiar spectral features in the cosmic neutrino spectrum observed by IceCube.}

\begin{document}

\maketitle

\section{Introduction}\label{sec:intro}
After more than 40 years from its theoretical prediction~\cite{Freedman:1973yd}, the COHERENT collaboration has recently measured Coherent Elastic Neutrino-Nucleus Scattering (\cevns)~\cite{COHERENT:2017ipa} using neutrinos produced at the Spallation Neutron Source (SNS) at the Oak Ridge National Laboratories, USA. The process is characterized by the small momentum exchange between neutrino and nucleus (below few tens of MeV) in such a way that the incoming neutrino interacts with the nucleus as a whole, rather than with the single nucleon or quark. The result is a huge enhancement in the \cevns cross section, proportional to the square of the number of nucleons. The interest in \cevns is two-fold: on the one hand, in the Standard Model (SM) context, it allows for the measurement of the sine square of the weak angle at low energy and for the measurement of poorly known nuclear properties like the neutron skin (see, for instance, Ref.~\cite{Cadeddu:2020lky}). On the other hand, assuming said nuclear properties can be extracted independently (for instance, by the PREX-II collaboration~\cite{PREX:2021umo}), \cevns becomes a powerful tool to probe New Physics (NP). In particular, we expect that in some region of parameter space the cross section will be enhanced, leading to potentially measurable results. In addition, \cevns is of paramount importance in dark matter direct detection experiments where it will limit the discovery potential (see~\cite{Strigari:2009bq} for the effect in a pure SM context, or~\cite{Harnik:2012ni,Bertuzzo:2017tuf,Gonzalez-Garcia:2018dep,Boehm:2018sux} for the effect of NP).

With the ``discovery'' phase behind our backs, we must now turn to a ``precision'' phase, aiming at increasing the statistics and precision of the measurements. To this end, various proposals have been put forward~\cite{Baxter:2019mcx,AristizabalSierra:2021uob, Fernandez-Moroni:2021nap}. The aim is to fully develop the \cevns physics case and overcome the experimental difficulty of detecting very small nuclear recoil energies. In this paper we are going to focus on two conceptually different proposals: (i) the directional low-pressure TPC detector $\nu$BDX-DRIFT~\cite{Snowden-Ifft:2018bde} using neutrinos produced in the Long Baseline Neutrino Facility (LBNF) at Fermilab, and (ii) a possible experiment to be installed at the European Spallation Source (ESS) in Lund, Sweden~\cite{Baxter:2019mcx}. The $\nu$BDX-DRIFT proposal is interesting because it employs neutrinos with higher energies with respect to those used at SNS and the directional sensitivity should allow for an improvement in the signal/background discrimination. On the other hand, the ESS proposal is interesting because it would allow to increase by approximately one order of magnitude the neutrino flux with respect to the one of the SNS. 
The ability of such experiments to contribute to the study of \cevns physics has been studied before~\cite{AristizabalSierra:2021uob,Baxter:2019mcx,Coloma:2020nhf,Chaves:2021pey}. Our purpose here is to extend such studies considering an interesting class of new physics models in which new light spin-1 particles are present. We will study three possibilities that have already been analyzed in connection with CE$\nu$NS: the universal $Z'$ model~\cite{Liao:2017uzy,Billard:2018jnl,Denton:2018xmq,Papoulias:2019txv,Papoulias:2019xaw}, the $B-L$ model~\cite{Billard:2018jnl,Han:2019zkz} and the $L_\mu -L_\tau$ model~\cite{Altmannshofer:2019zhy,Abdullah:2018ykz}. For each model we will compute the reach of both the proposals mentioned above, comparing them with current exclusions coming from COHERENT and other experiments.

The paper is organized as follows. In Sec.~\ref{sec:models} we present in some detail the light $Z'$ models we will consider. In Sec.~\ref{sec:experiments} we discuss the $\nu$BDX-DRIFT and ESS proposals, while we devote Sec.~\ref{sec:sensitivity} to the phenomenological study of the models of interest. Finally, we present our conclusions in Sec.~\ref{sec:conclusions}.

\section{Light vector mediator models}\label{sec:models}
Models with light new abelian gauge bosons have received increasing attention over the last few years as an effective way to enhance CE$\nu$NS with respect to the SM~\cite{Liao:2017uzy,Billard:2018jnl,Denton:2018xmq,Papoulias:2019txv,Papoulias:2019xaw,Han:2019zkz,Altmannshofer:2019zhy,Abdullah:2018ykz}. For our purposes, the relevant interaction is of the type
\be\label{eq:Lint}
{\cal L}_{\rm int} = g_{Z'} \, Z^\prime_\mu \sum_f Q_f' \bar{f} \gamma^\mu f,
\ee
where $f$ are SM fermions.\footnote{We assume for simplicity that there is no tree-level kinetic mixing between the hypercharge boson and the $Z'$.} The charges $Q_f'$ vary from model to model and will be given later for the cases of interest. Since we are interested in CE$\nu$NS, it is essential that the $Z'$ has a non-vanishing coupling to neutrino states. We will consider only left handed neutrinos: possible right handed neutrinos, if present in the model, are supposed to be sufficiently heavy to be neglected. Notice that in Eq.~\eqref{eq:Lint} we write only the NP interactions. Needless to say, the usual SM neutral interactions are also present and must be included in the computation of the scattering cross section. The total differential cross-section for the \cevns reaction $\nu T \to \nu T$ over a nuclear target $T$ reads
\be\label{eq:cevns_cs}
\frac{d\sigma}{d E_r} = \frac{G_F^2\, m_T}{\pi}\left(1- \frac{m_T E_r}{2 E_\nu^2} \right) Q^2(E_r),
\ee
where $G_F$ is the Fermi constant, $m_T$ is the nucleus mass, $E_r$ the nuclear recoil energy and $E_\nu$ the incident neutrino energy. Finally, $Q^2(E_r)$ contains the coherent enhancement factor and depends on the $Z$ and $Z'$ bosons interactions with fermions:
\be\label{eq:total_charge}
Q(E_r) = Z g_p F_p(E_r) + (A-Z) g_n F_n(E_r) .
\ee
In the previous expression, $Z$ and $A$ are, respectively, the proton and mass number of the target nucleus; $F_{p,n}(E_r)$ are the proton and neutron nuclear form factors (see below); finally, $g_{p,n}$ are the total couplings to protons and neutrons. In the low energy limit we are interested in, $g_{p,n}$ can be computed simply by summing over the individual quark contributions $g_q$ inside the nucleon. Each quark contributes
\be\label{eq:quark_coupling}
g_q = \left(T_{3L}^q - 2 s_W^2 Q_q\right) + \frac{g_{Z'}^2\, Q_\nu'\, Q_q'}{\sqrt{2} G_F \left( |\mathbf{q}|^2 + m_{Z'}^2 \right) }, 
\ee
with $|\mathbf{q}| = \sqrt{2 m_T E_r}$ the modulus of the 3-momentum exchanged in the reaction. The first contribution is due to the SM $Z$ exchange ($T_{3L}^q$ is the $SU(2)_L$ quantum number of the quark $q$, $Q_q$ its electric charge and $s_W$ the sine of the weak angle) while the second one is due to the $Z'$ exchange. In the $|\mathbf{q}| \ll m_{Z'}$ limit the second term recovers the neutrino Non-Standard-Interaction (NSI) parameter $\epsilon^f_{\nu\nu}$. Summing the individual quark contributions shown in Eq.~\eqref{eq:quark_coupling} we recover a well-known result: the SM contribution for the proton is very small, while for the neutron it is larger and negative. This means that there will be an enhancement of the cross section in the extended model, should the NP contribution be negative, due to constructive interference with the SM one. Likewise, there will be some cancellation between the SM and NP contributions should the latter be positive, strongly suppressing the total cross-section for some values of the NP parameters. Furthermore, in this case, we may define a region in parameter space where the NP contribution approaches twice the SM one, such that $Q(E_R) = -Q_{SM}(E_R)$, with the left hand side defined as in Eq.~\eqref{eq:total_charge} with the couplings given by eq.~\eqref{eq:quark_coupling} and $Q_{SM}$ following the same definition with $g_{Z'}=0$. Since the cross section is proportional to $Q(E_R)^2$, in this region the total number of events approaches the SM value, and the bounds may vanish. We will get back to this feature in Sec.~\ref{sec:sensitivity}.

We now introduce the three light $Z'$ models we will consider:
\begin{description}
\item[$Z'_{\rm universal}$:] the new light gauge boson couples to all fermions with the same strength, i.e. $Q_q' = Q_\nu' = 1$;
\item[$Z'_{B-L}$:] the light $Z'$ couples to the (anomaly free) current of $U(1)_{B-L}$ with $B$ and $L$ the baryon and lepton numbers, respectively. As a consequence, quarks have charge $Q_q' = 1/3$ and neutrinos have charge $Q_\nu' = -1$;
\item[$Z'_{\mu-\tau}$:] in this case, the $Z'$ couples to the $U(1)_{L_\mu - L_\tau}$ current, i.e. $Q_{\nu_e}' = 0 = Q_q'$ and $Q_{\nu_\mu}' = 1 = - Q_{\nu_\tau}'$. Since the quarks do not interact with the $Z'$, the process proceeds via a muon and tau loop generating a kinetic mixing between the photon and the $Z'$. Eq.~\eqref{eq:quark_coupling} must be replaced by~\cite{Ballett:2019xoj,Corona:2022wlb}
\begin{align}\begin{aligned}
g_q^{\mu-\tau} & = \left(T_{3L}^q - 2 s_W^2 Q_q\right) \\ 
& \qquad + \frac{2\, \alpha_{EM}\, g_{Z'}^2\, Q_q}{\sqrt{2} \pi\, G_F \left( |\mathbf{q}|^2 + m_{Z'}^2 \right)} \int_0^1 dx\, x (1-x) \log\left[\frac{m_\tau^2 + |\bm{q}|^2 x(1-x)}{m_\mu^2 + |\bm{q}|^2 x(1-x)}\right] .
\end{aligned}\end{align}
In this model, clearly, only protons will contribute to the total cross section at leading order in the exchanged momentum.
\end{description}

We conclude this section with an explicit expression for the nuclear form factors. For simplicity, we take equal form factors for protons and neutrons and adopt the Helm parametrization \cite{HelmFF}
\be
F_{p,n}(E_r) = \frac{3 j_1(|\mathbf{q}|\, R^{p,n}_0)}{|\mathbf{q}|\, R^{p,n}_0}e^{-|\mathbf{q}|^2 s^2/2}, ~~~ R_0^{p,n} = \left(\frac{5}{3} \langle r_{p,n}^2 \rangle - 5 s^2 \right)^{1/2}, 
\ee
with $j_1$ the spherical Bessel function of the first kind, $s = 0.9$ fm \cite{Lewin:1995rx}, $|\mathbf{q}|\,$ as previously defined below \eq{eq:quark_coupling} and $\langle r^2_p \rangle = \langle r^2_n \rangle$ for the different targets taken from~\cite{Angeli:2013epw}. We show the values used in this work in Table~\ref{tab:nuclear_radii}.

\begin{table}[t!]
    \centering
    \begin{tabular}{|c|c|c|c|c|c|c|c|c|c|}
    \hline
    & $^{133}$Cs & $^{127}$I & $^{28}$Si & $^{132}$Xe & $^{72}$Ge & $^{40}$Ar & $^{12}$C & $^{19}$F & $^{32}$S \rule{0pt}{2ex}  \rule[-1.ex]{0pt}{0pt}\\
    \hline
    $\langle r_p^2\rangle^{1/2}$ [fm] & 4.80 & 4.75 & 3.12 & 4.79 & 4.06 & 3.43 & 2.47 & 2.90 & 3.26 \rule{0pt}{2.5ex}  \rule[-2.0ex]{0pt}{0pt}\\
    \hline
    \end{tabular}
    \caption{Isotopes and respective nuclear radius values, taken from \cite{Angeli:2013epw}.}
    \label{tab:nuclear_radii}
\end{table}

\section{Future facilities}\label{sec:experiments}
We devote this section to the description of the future proposals already mentioned in Sec.~\ref{sec:intro}. In both cases, we will compute the rate of recoil events, given by
\begin{equation}\label{eq:event_rate}
   R = N_T \int_{E_r^\text{min}}^{E_r^\text{max}}\int_{E_\nu^\text{min}}^{E_\nu^\text{max}}dE_rdE_\nu\frac{d\sigma}{dE_r}\frac{d\phi}{dE_\nu},
\end{equation}
as a function of the parameters of the models. 
In the previous equation, $N_T$ is the number of nuclei in the target, $d\sigma/dE_r$ is the neutrino-nucleus cross section of Eq.~\eqref{eq:cevns_cs} and $d\phi/d E_\nu$ the differential incident neutrino flux. 
We will describe in Secs.~\ref{sec:nuBDX} and~\ref{sec:ESS} how to compute these quantities. 
The recoil energy, $E_r$, and the neutrino energy, $E_\nu$, are related by
\begin{equation}\label{eq:er_enu_relation}
    E_\nu = \sqrt{\frac{E_r m_T}{2}}.
\end{equation}

We will, again, describe in Secs.~\ref{sec:nuBDX} and~\ref{sec:ESS} how the extrema of integration in Eq.~\eqref{eq:event_rate} are determined. 

\subsection{$\nu$BDX-DRIFT}\label{sec:nuBDX}

The $\nu$BDX-DRIFT proposal is based on the Beam Dump eXperiment - Directional Recoil Identification From Tracks (BDX-DRIFT), an experimental proposal employing a negative ion drift TPCs to achieve directional resolution of low threshold recoil events~\cite{PhysRevD.61.101301}. Although originally conceived as a Dark Matter detector, a recent paper has explored its repurposing as a CE$\nu$NS experiment~\cite{AristizabalSierra:2021uob}, to be placed in the planned Near Detector Facility of the DUNE complex at Fermilab.

A BDX-DRIFT detector consists of a collection of 1m $\times$ 1m $\times$ 1m box units placed behind a beam dump. In the configuration discussed in Ref.~\cite{1809.06809}, the detector is filled with a mixture of 40 Torr CS$_2$ + 1 Torr O$_2$. Two readout planes with a central cathode between them are to be placed along the beam direction, configuring two drift volumes. Each readout plane would be composed of several anode wires connected to a gain element, to which the ionized gas molecules can release its electrons for normal avalanche to occur. The distance between wires is related to the minimum track resolution length and sets a limit on the lower threshold resolution. The gaseous elements are chosen because of the following properties: CS$_2$, having a strong electronegative nature, allows for the ionization tracks produced by recoil events to be transported through the ions themselves, largely preserving the track shape~\cite{1301.7145}; O$_2$, on the other hand, allows to measure the distance (in the beam direction) between the initial event and the readout planes through the production of additional ions drifting at different speeds in the uniform electric field~\cite{1301.7145}. The overall result is a careful 3D fiducialization of the active detector volume that allows for a very good background rejection, in particular with respect to non-beam related sources. Previous runs of a CS$_2$ and O$_2$ DRIFT detector have indeed shown very clean results for Dark Matter searches~\cite{DRIFT:2014bny}.

The $\nu$BDX-DRIFT proposal~\cite{AristizabalSierra:2021uob} takes advantage of the BDX-DRIFT detector setup described above to achieve a good background mitigation and low detection thresholds. In a CE$\nu$NS event, most of the recoil events off Sulfur nuclei (with recoil energies up to tens of keVs) would scatter in a direction lying, at most, at 1 degree from the beam line, allowing for careful exclusion of most types of background-induced signals. Beam-related backgrounds as neutrino-induced neutrons or inelastic neutrino processes, on the other hand, could potentially induce scattering events similar to CE$\nu$NS events. Their contribution has however been computed in~\cite{AristizabalSierra:2021uob} and shown to be largely subdominant for $\nu$BDX-DRIFT. 

While previous deployments of DRIFT detectors worked mostly with a fixed 40 (+1) Torr configuration, it has been suggested in~\cite{AristizabalSierra:2021uob} that other possibilities could maximize the CE$\nu$NS yield, and other gases have also been considered and tested as viable alternatives to CS$_2$ (eg. CF$_4$ in~\cite{1301.7145}). These different configurations change the expected event number modifying the number of targets and the recoil energy threshold. Considering a detector with a fixed active volume, working at room temperature, and taking a perfect gas approximation, the number of targets $N_T$ is determined by
\begin{equation}\label{eq:NT_bdx}
    N_T = \rho(P)\frac{N_A}{m_{\text{molar}}}V_{\text{det}}, \hspace{1cm} \rho = 5.5\times10^{-5}\times\left(\frac{m_\text{molar}}{\text{g/mol}}\right)\left(\frac{P}{\text{Torr}}\right)\frac{\mathrm{kg}}{\mathrm{m}^3},
\end{equation}
where $N_A$ is the Avogadro number, $V_{\text{det}}$ is the detector volume, $m_{\text{molar}}$ is the molar mass of the gas in the detector and $P$ is the pressure. Turning to the dependence of $E_r^{\rm min}$ on the gas pressure and chemical composition, we have
\begin{equation}\label{eq:Emin_bdx}
E_r^{\mathrm{min}} = E_{\mathrm{th}}(\text{Nuc}_i) = f_i\left(\frac{P}{40\, \text{Torr}}\right)\text{ keV},    
\end{equation}
with $f_i=\{7.5,\,20\}$ for $\text{Nuc}_i=\{C,\,S\}$, respectively~\cite{DRIFT:2014bny}. Eq.~\eqref{eq:Emin_bdx} takes into account that the lower threshold for the recoil energy depends on the length of the tracks that can be resolved when they reach the readout, with lighter nuclei producing longer tracks and thus having a lower threshold. In addition, using the Bethe-Bloch equation, the energy loss scales with the density of the medium, limiting track length for a given recoil energy. Since we are working in the ideal gas approximation, the density dependence can be converted into a pressure dependence using Eq.~\eqref{eq:NT_bdx}. Comparing Eq.~\eqref{eq:NT_bdx} and~\eqref{eq:Emin_bdx} we see that larger pressures lead to a larger number of targets, but at the same time increase the minimum recoil energy. There is thus a trade-off that must be considered in searching for the ideal pressure configuration given the chemical composition of the gas inside the DRIFT chamber. 
\begin{figure}
    \centering
    \includegraphics[width=0.8\textwidth]{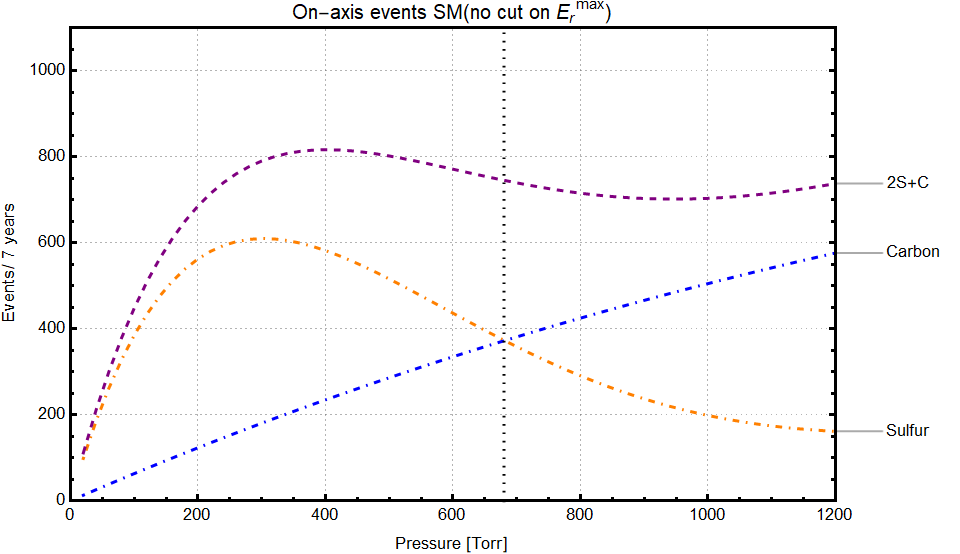}
    \caption{Expected number of CE$\nu$NS events in $\nu$BDX-DRIFT as a function of the pressure configuration (in Torr) for a data-taking period of 7 years. The vertical grey line shows the pressure value where the Carbon contribution overtakes the Sulfur one. The number of events is computed in the SM and, for illustrative purposes, we do not impose any cut on the maximum recoil energy.}
    \label{fig:bdx_SMevents}
\end{figure}

In order to finally compute the event rate of Eq.~\eqref{eq:event_rate}, two ingredients are missing: the incident neutrino flux and the value of $E_r^{\rm max}$. We consider the incident neutrino flux for the LBNF Beam Dump described in the DUNE Experiment's Technical Report (see~\cite{DUNE:2020ypp}, Figure 4.9). Due to the background considerations mentioned above, we consider the on-axis configuration for the beam, in $\nu$ mode. The flux composition is well approximated by a pure $\nu_\mu$ beam up to the position where the $\nu$BDX-DRIFT detector would be placed, in the Near Detector Facility. It peaks between 2 and 3 GeV, with a tail extending all the way up to 5 GeV. This motivates the choice $E_\nu^{\text{max}} = 5$ GeV for the maximum neutrino energy. As a consequence, the recoil energy cutoff is $E_r^{\text{max}} = 2(E_\nu^{\text{max}})^2/m_T$ and depends on the nuclear species through the target's mass $m_T$. Each of the Carbon and Sulfur nuclei can scatter independently, so it is crucial to understand whether these two types of events can be distinguished, and how any energy cut must be imposed. Given the $\nu$BDX-DRIFT's capabilities, those can only be done via track length cuts, which cannot differentiate between nuclear species in most of the phase space, generating ambiguity on the recoil energy of the event. This is not an issue, since we are only using the total rate. This can be computed simply as a weighted sum of the individual rates for each species, with coefficients determined by stoichiometry. The oxygen contribution can be neglected, being present only in very small quantities, leading to
\begin{equation}
    R_{\text{Total}} = R(C) + 2 R(S).
\end{equation}
Each of the rates is computed using Eq.~\eqref{eq:event_rate}. As mentioned above, we do not impose cuts on the maximum incoming neutrino energy, using the whole available flux, since the background has been computed up to $5$ GeV in Ref.~\cite{AristizabalSierra:2021uob} and confirmed to be subdominant. As for the maximum recoil energy, we take $E_r^{\text{max}} = 3192$ keV, corresponding to the Carbon limit for 5 GeV maximum neutrino energy. This range of recoil energy includes completely the interval of recoil energies available to Sulfur events. This is experimentally justified since the recoil energy cut corresponds to a track length cut and, in the absence of a clear correspondence between recoil energy and track length for each nucleus, our choice does not discard any Carbon or Sulfur event.  The predicted event count for the Standard Model, considering a science run of 7 years, is shown in Figure \ref{fig:bdx_SMevents}, plotted against the pressure configuration of the detector.

\subsection{European Spallation Source}\label{sec:ESS}

The European Spallation Source (ESS) is a multi-disciplinary research facility currently under construction on the outskirts of Lund, Sweden. Although it has been designed to provide the world's most powerful neutron source, its $\pi$DAR (``pion Decay-At-Rest'') setup also yields a neutrino flux which can be used for other kinds of experiments. As a result, some groups have suggested a CEvNS oriented detector program to be incorporated in the ESS future facilities~\cite{esteban2021european}. Several detector technologies have been proposed~\cite{Baxter:2019mcx}: a Cryogenic (77 K) undoped CsI scintillator array; Low-background Silicon-based CCD arrays with single-electron threshold; a High-pressure gaseous Xenon TPC; a p-type point contact Germanium detector; Liquid Argon and C$_3$F$_8$-based bubble chambers. The corresponding threshold energy and energy resolution for each detector are listed in Table~\ref{tab:detectors}. All the values are given in keVnr, representing the underlying true nuclear recoil energy for each event. 

\begin{table*}[t!]
  \renewcommand{\arraystretch}{1.4} \centering
{\scriptsize
 \begin{tabular}{|@{\hspace*{2pt}}c@{\hspace*{2pt}}|@{\hspace*{2pt}}c@{\hspace*{2pt}}|@{\hspace*{2pt}}c@{\hspace*{2pt}}|@{\hspace*{2pt}}c@{\hspace*{2pt}}|@{\hspace*{2pt}}c@{\hspace*{0pt}}|@{\hspace*{2pt}}c@{\hspace*{2pt}}|@{\hspace*{2pt}}c@{\hspace*{2pt}}|}
 \hline
 	Detector Technology & Target & Mass [kg]
                 & $E_r^{\mathrm{min}}$ [keV$_{\mathrm{nr}}$]& $\frac{\Delta E}{E}\bigg|_{E_{\mathrm{th}}}$
                (\%) & E$_r^{\mathrm{max}}$ [keV$_{\mathrm{nr}}$] & background [day$^{-1}$] 	\rule{0pt}{4ex}  \rule[-3.5ex]{0pt}{0pt} \\ 
                \hline
                Cryogenic scintillator & CsI &
                22.5 &  1 & 30 &
                46.1 & 406 \\
                Charge-coupled device & Si & 1 &
                0.16 & 60 & { 212.9} & 8.5\\
                High-pressure gaseous TPC &
                Xe & 20 & 0.9 & 40 & 45.6 & 357.6 \\
                p-type point contact HPGe
                & Ge & 7 & 0.6 & 15 & 78.9 & 329 \\
                Scintillating bubble chamber & Ar &
                10 &  0.1 & 40 & 150.0 & $4\times 10^{-2}$ \\
                Standard bubble chamber & C$_{3}$F$_{8}$ & 10
                & 2 & 40 & 329.6 & $4\times 10^{-2}$
                \\ 
                \hline
  \end{tabular}
}  
	\caption{\label{tab:detectors} Summary of properties for the detectors possibly employed at the ESS: target, mass, recoil energy threshold, width for the smearing and maximum recoil energy, steady-state background. The steady-state background include the $4\times10^{-2}$ reduction by the ESS duty factor.}
\end{table*}

Since the CE$\nu$NS cross-section enhancement depends on the mass number of the target nuclei, ideally several of the technologies listed above may be used in order to test with a good precision the SM prediction. This strategy also has the advantage of potentially lifting possible blind spots in the analysis of NP models~\cite{Chaves:2021pey}.

As already mentioned, the ESS CE$\nu$NS program would operate with the neutrino flux generated by a pion decay-at-rest ($\pi$DAR) setup, in which accelerated protons collide with a Tungsten target inside a shielded monolith, producing neutrons and pions. Pions are stopped by the shielding and decay into neutrinos. The neutrino flux has two components: (i) a prompt one, generated directly by pion decay, composed by monochromatic $\nu_\mu$ with energy $\frac{m_\pi^2-m_\mu^2}{2m_\pi}\simeq 29.7$ MeV; (ii) a delayed one, generated by muons decaying in-flight, consisting of a $\nu_e - \bar{\nu}_\mu$ mixture with energy up to $E_{\nu_e,\Bar{\nu}_{\mu}}<\frac{m_\mu}{2}\simeq 52.8$ MeV. In our analysis we will use the following analytic form for the fluxes:
\begin{equation}\label{eq:ESS_flux}
\begin{split}
    \frac{d\phi(\nu_\mu)}{dE_\nu} & = \eta\, \delta\left(E_\nu-\frac{m_\pi^2-m_\mu^2}{2m_\pi}\right), \\
    \frac{d\phi(\Bar{\nu}_\mu)}{dE_\nu} & = \eta \frac{64}{m_\mu}\left[\left(\frac{E_\nu}{m_\mu}\right)^2\left(\frac{3}{4}-\frac{E_\nu}{m_\pi}\right)\right]\theta\left(E_\nu-\frac{m_\mu}{2}\right), \\
    \frac{d\phi(\nu_e)}{dE_\nu} & = \eta \frac{192}{m_\mu}\left[\left(\frac{E_\nu}{m_\mu}\right)^2\left(\frac{1}{2}-\frac{E_\nu}{m_\pi}\right)\right]\theta\left(E_\nu-\frac{m_\mu}{2}\right),
\end{split}\end{equation}
with $\theta(x)$ the Heaviside function. The normalization factor $\eta$ is given by
\begin{equation}
    \eta = \phi_{\nu/POT}\frac{N_{POT}}{4\pi L^2}, 
\end{equation}
with $\phi_{\nu/POT}$ the number of neutrinos produced per proton-on-target, $N_{POT}$ the number of protons-on-target (per time), and $L$ the distance from the monolith source to the detector. Following~\cite{Baxter:2019mcx}, we take $\phi_{\nu/POT} =0.3$, $N_{POT} = 2\times 10^{23}$/year and $L = 20$ m, respectively. As was done for $\nu$BDX-DRIFT, we consider the whole available range of neutrino energies, with $E_\nu^{\text{max}}= 52.8$ MeV and $E_r^{\text{max}}$ computed according to Eq.~\ref{eq:er_enu_relation}. We list in Tab.~\ref{tab:detectors} the maximum recoil energy for each of the possible detectors.

\section{Physics sensitivity}\label{sec:sensitivity}

In what follows we derive the sensitivities on the mass $m_{Z'}$ and coupling $g_{Z'}$ of a light mediator that couples with the SM particles, as described in Sec. \ref{sec:models} and \ref{sec:experiments}.

The sensitivity to a new physics model characterized by a coupling $g_{Z'}$ and a mass $m_{Z'}$ is obtained by a binned $\chi^2$:
\begin{eqnarray}
\chi^2(g_{Z'}, m_{Z'}, \xi_S, \xi_B) &=& \sum_i 2\biggl[\xi_S N_i(g_{Z'},m_{Z'}) + \xi_B N_{\mathrm{B},i} \nonumber\\
&-& (N_{\mathrm{SM},i} + N_{\mathrm{B},i})\left(1 - \ln\left( \frac{N_{\mathrm{SM},i} + N_{\mathrm{B},i}}{\xi_S N_i(g_{Z'},m_{Z'}) + \xi_B N_{\mathrm{B},i}}\right)  \right)   \biggr] \nonumber \\
&+& \left(\frac{1-\xi_S}{\sigma_S}\right)^2 + \left(\frac{1-\xi_B}{\sigma_B}\right)^2,
\end{eqnarray}
where $N_i$ and $N_{\mathrm{B},i}$ are the event and background rate in the $i$-th energy bin, $\xi_S$ and $\xi_B$ are the normalization factors for signal and background respectively, and $N_{\mathrm{SM}}$ is the expected SM event rate. The last two terms take into account the systematic uncertainties, exploiting the pull method, where the signal and background normalization uncertainties are denoted by $\sigma_S$ and $\sigma_B$, respectively. 

For the $\nu$BDX-DRIFT experiment \cite{AristizabalSierra:2021uob}, we employ a simple single bin $\chi^2$, due to the difficulty for the $\nu$BDX-DRIFT experiment to do a spectral analysis, and we will show the results for a CS$_2$ target detector with a volume of $10$ m$^3$ and seven years data taking. We assume that the background is a fraction $f=25\%$ of the expected number of events. The signal normalization uncertainty is due mostly to our ignorance of the nuclear form factor and the neutrino flux, taken to be $\sim10\%$ and summed in quadrature. We furthermore assume that the systematic uncertainty on the background is of $\sim 1\%$.

For the experiments at the ESS we assume that the targets are at $20$ m, the detector signal acceptance is $\epsilon=0.8$ above threshold and the running time is limited to $3$ years \cite{Baxter:2019mcx}. The detector properties, energy thresholds  and expected background rates are taken from Table 1 of Ref. \cite{Baxter:2019mcx} and listed in Table \ref{tab:detectors}. 
The sensitivity is computed for all detectors, with the exception of the Ar and C$_3$F$_8$ bubble chambers, binning the energy rates in such a way that, for each bin, the bin size is twice the energy resolution at its center. Furthermore, we apply a gaussian energy smearing with a width $\sigma(E_r) = \sigma_0 \sqrt{E_r/E_r^{\mathrm{min}}}$, where $\sigma_0$ is the energy resolution at the energy threshold $E_r^{\mathrm{min}}$. For the Ar and C$_3$F$_8$ detectors, on the other hand, we use the total unbinned rate. Finally, following Ref. \cite{Baxter:2019mcx}, we assume the signal and background normalization uncertainties to be $\sim 10\%$ and $\sim1\%$, respectively. 

{Although a detailed analysis is beyond the scope of this work, modifying the binning of our analyses may improve the bounds. This is particularly true at small masses, where the distribution peaks at low energy. Defining different region of interest depending on the $Z'$ mass will be crucial once the data will be available.}

\subsection{Sensitivity for the universal $Z'$ model}

\begin{figure}[t!]
  \begin{center}
    \includegraphics[width=0.49\textwidth]{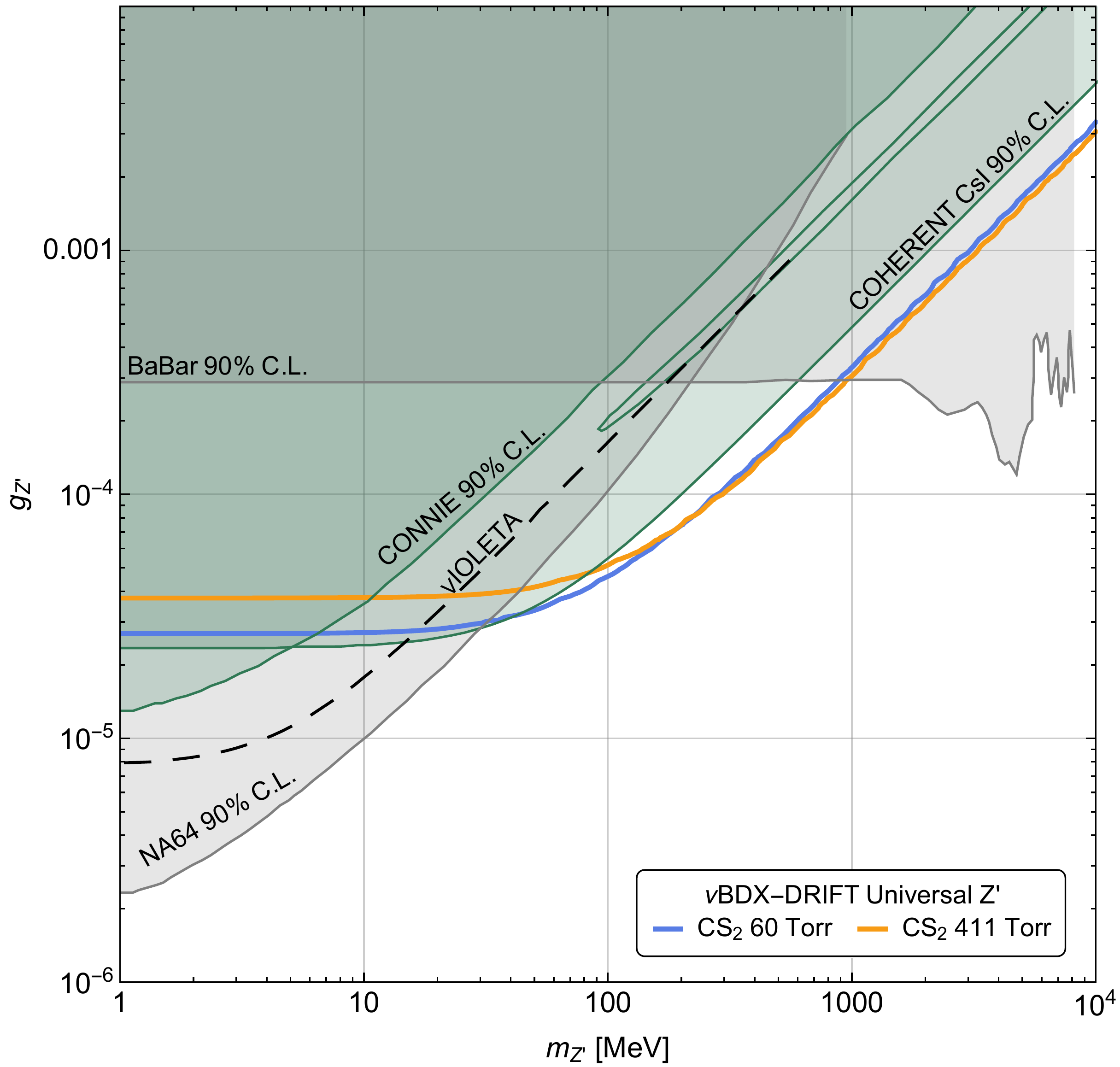}
    \includegraphics[width=0.49\textwidth]{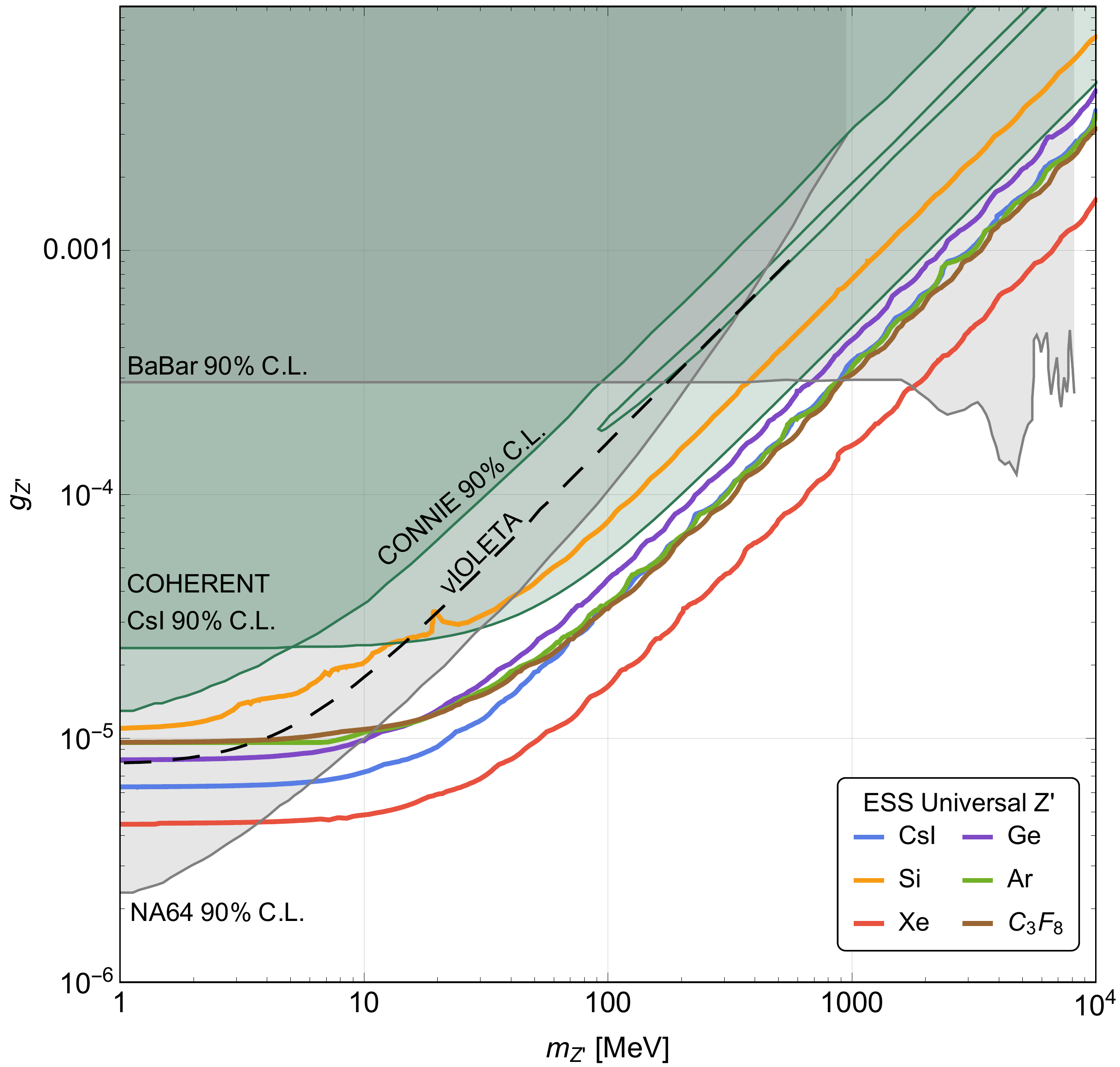}
    \caption{Future sensitivity at $90\%$ C.L. in the $m_{Z'} - g_{Z'}$ plane for the universal $Z'$ model for the $\nu$BDX-DRIFT detector (left) exploiting CS$_2$ at two different pressures, and the various detectors described in the text at the ESS (right). The dark green regions are excluded at $90\%$ C.L. by CONNIE \cite{CONNIE:2019xid} and COHERENT \cite{Cadeddu:2020nbr} assuming dominant $Z'$ decays to SM states. The dashed black curve shows the sensitivity of $\nu$IOLETA \cite{Fernandez-Moroni:2021nap}. The gray shaded areas are excluded at $90\%$ C.L. by BaBar \cite{BaBar:2017tiz} and NA64 \cite{NA64:2017vtt,Banerjee:2019pds} assuming that the $Z'$ decays dominantly in invisible dark sector states.    \label{fig:UniversalZp}}
  \end{center}
\end{figure}

Figure \ref{fig:UniversalZp} shows the $90\%$ C.L. limits obtained using the $\nu$BDX-DRIFT detector at the LBNF (left) and the various detectors exploiting the ESS neutrino source (right) for the universal $Z'$ model. The left panel shows sensitivity curves for two different values of the pressure of the CS$_2$ gas used: the blue curve simulates a detector with a pressure of 60 Torr, while the orange one exploits $411$ Torr, the pressure that Ref.~\cite{AristizabalSierra:2021uob} shows to give the largest rate for the SM case. 
{For masses $m_{Z'} \gtrsim 100$ MeV the two curves are superimposed, while for smaller values of the $Z'$ mass the limits are similar but not completely equal, with the $P=60$ Torr case being slightly more stringent. This is due to the dependence of Eq.~\eqref{eq:quark_coupling} on the parameters: for $m_{Z'} \gtrsim 100$ MeV the term proportional to $|\bm{q}|^2$ is subdominant, explaining the universal behavior of the two curves. On the contrary, for smaller $Z'$ masses, the term $|\bm{q}|^2$ is important and its dependence on the target mass justifies the difference between the blue and orange curves.
}
{We also stress that, although} the mass of the detector increases linearly with the pressure, see Eq. \eqref{eq:NT_bdx}, increasing the number of events, a larger pressure increases also the energy threshold, Eq. \eqref{eq:Emin_bdx}, dramatically cutting out regions in energy where the rate is larger.  

The right panel, on the other hand, shows the sensitivity for the six detectors under consideration exploiting CsI (blue), Ge (purple), Si (orange), Ar (green), Xe (red) and C$_3$F$_8$ (brown). Here we notice that the stronger sensitivity comes from the detectors with the largest value of the atomic mass number A (Xenon and CsI), as it would be naively expected since they give the largest number of events. An exception is the detector with C$_3$F$_8$. This is due to the fact that its energy threshold is the largest among all the detectors. 

In both panels, there is a thin diagonal strip, as shown in \cite{Cadeddu:2020nbr}, that is not excluded because it corresponds to values of the couplings and mass for which the SM and NP contributions combine to approximately recover the expected number of events in the SM (see the discussion after Eq.~\eqref{eq:quark_coupling}). This region can be probed by the interplay between the sensitivity results for the different target materials of the detectors at the LBNF and ESS facilities. These same considerations apply for the $L_\mu-L_\tau$ model. 

For both panels, the dark green shaded regions are already excluded at $90\%$ C.L. by CONNIE \cite{CONNIE:2019xid} and COHERENT \cite{Cadeddu:2020nbr}. 
The black dashed curve furthermore shows the sensitivity that another proposed experiment, the Neutrino Interaction Observation with a Low Energy Threshold Array ($\nu$IOLETA), can reach for the universal $Z'$ model \cite{violeta,neutrino-2020-poster-1,neutrino-2020-poster-2,neutrino-2020-poster-3,Fernandez-Moroni:2020yyl,Fernandez-Moroni:2021nap}. These searches assume that the $Z'$ decays most of the time to SM states. If the $Z'$ decays predominantly to invisible dark sector particles, existing searches from the BaBar \cite{BaBar:2017tiz} and the NA64 \cite{NA64:2017vtt,Banerjee:2019pds} experiments apply and exclude the gray region. 

\subsection{Sensitivity for the $B-L$ model}

\begin{figure}[t!]
  \begin{center}
    \includegraphics[width=0.49\textwidth]{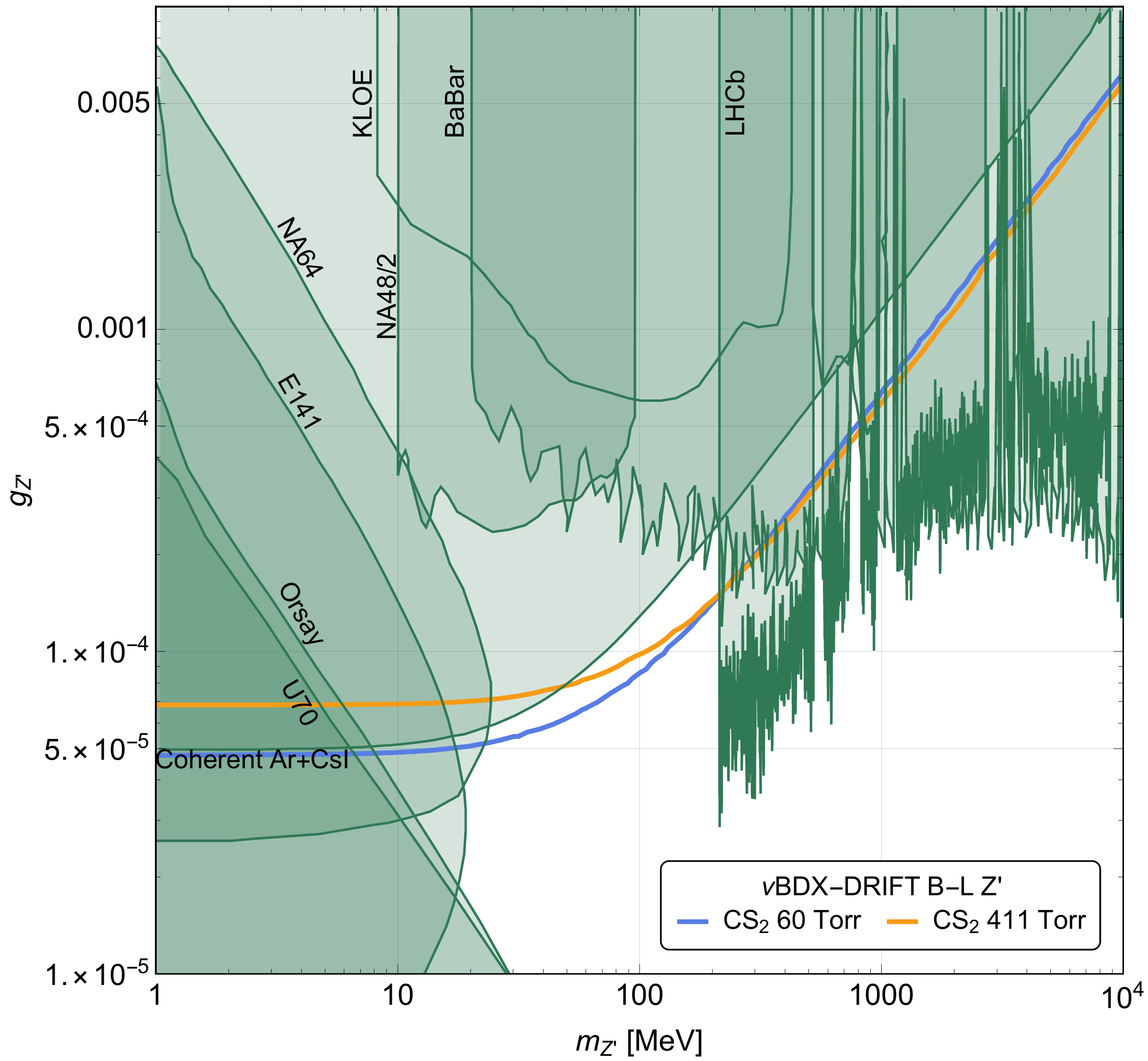}
    \includegraphics[width=0.49\textwidth]{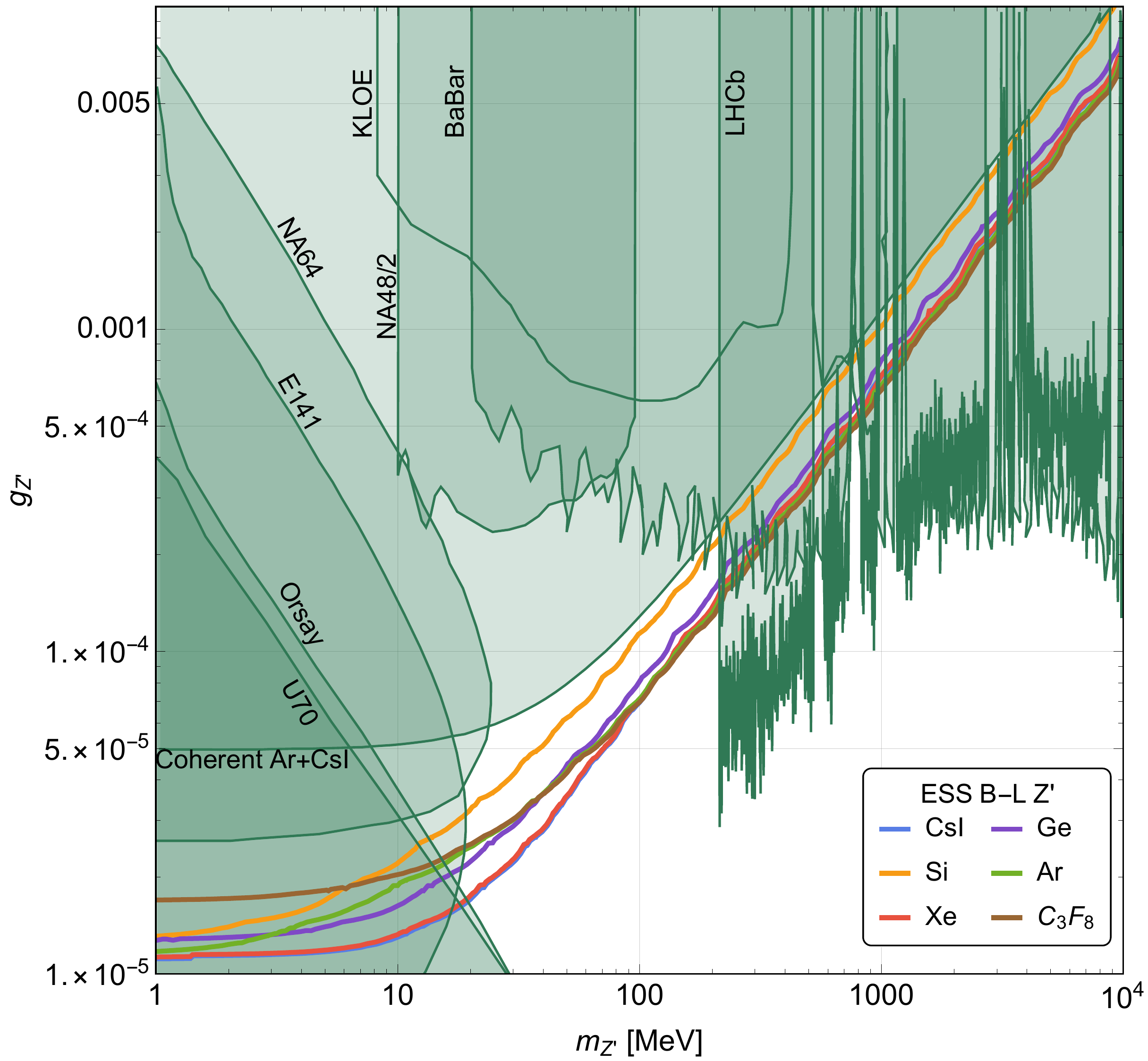}
    \caption{Future sensitivity at $90\%$ C.L. in the $m_{Z'} - g_{Z'}$ plane for the $B-L$ model for the $\nu$BDX-DRIFT detector (left) exploiting CS$_2$ at two different pressures, and the various detectors described in the text at the ESS (right). The dark green regions are excluded at $90\%$ C.L. by NA64 \cite{NA64:2019auh}, E141 \cite{PhysRevLett.59.755}, Orsay \cite{Davier:1989wz}, U70 \cite{Blumlein:2011mv}, COHERENT \cite{Cadeddu:2020nbr}, NA48/2 \cite{NA482:2015wmo}, KLOE \cite{KLOE-2:2011hhj,KLOE-2:2012lii}, BaBar \cite{BaBar:2014zli} and LHCb \cite{LHCb:2019vmc} assuming dominant $Z'$ decays to SM states. \label{fig:B-LZp}}
  \end{center}
\end{figure}

We show in Figure \ref{fig:B-LZp} the 90$\%$ C.L. sensitivity obtained for the $B-L$ model. The dark green shaded area is excluded by existing experimental searches assuming dominant $Z'$ decays into SM particles, from fixed target experiments (NA64 \cite{NA64:2019auh}, E141 \cite{PhysRevLett.59.755}, Orsay \cite{Davier:1989wz} and U70 \cite{Blumlein:2011mv}), COHERENT \cite{Cadeddu:2020nbr}, KLOE \cite{KLOE-2:2011hhj,KLOE-2:2012lii} and NA48/2 \cite{NA482:2015wmo}. Furthermore, as shown in Ref.~\cite{Ilten:2018crw}, it is possible to reinterpret the BaBar \cite{BaBar:2014zli} and LHCb \cite{LHCb:2019vmc} bounds on dark photons and adapt them to the $B-L$ case.

The panel on the left of Figure \ref{fig:B-LZp} shows the sensitivity for the $\nu$BDX-DRIFT detector filled with CS$_2$ at 60 Torr (blue) and 411 Torr (orange). This detector will be able to slightly increase the sensitivity of current experiments and will help improve the coverage between accelerator and fixed target experiments. The 90$\%$ C.L. sensitivity curves for the experiments at the ESS are shown in the right panel. These detectors, and in particular the one exploiting Xe, will be able to probe a good part of the parameter space that currently is not probed by COHERENT, collider or fixed target experiments.

\begin{figure}[t!]
  \begin{center}
    \includegraphics[width=0.49\textwidth]{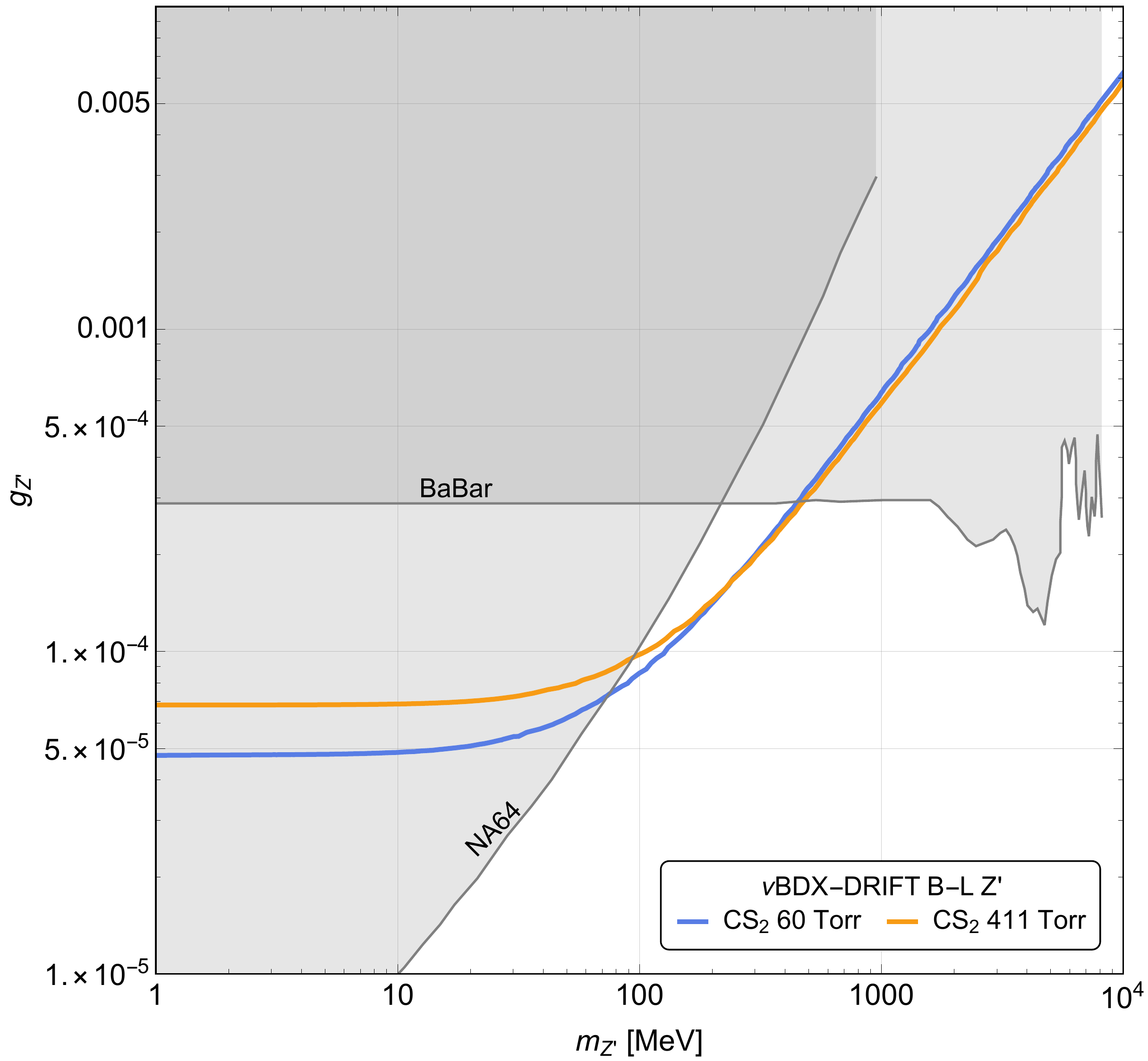}
    \includegraphics[width=0.49\textwidth]{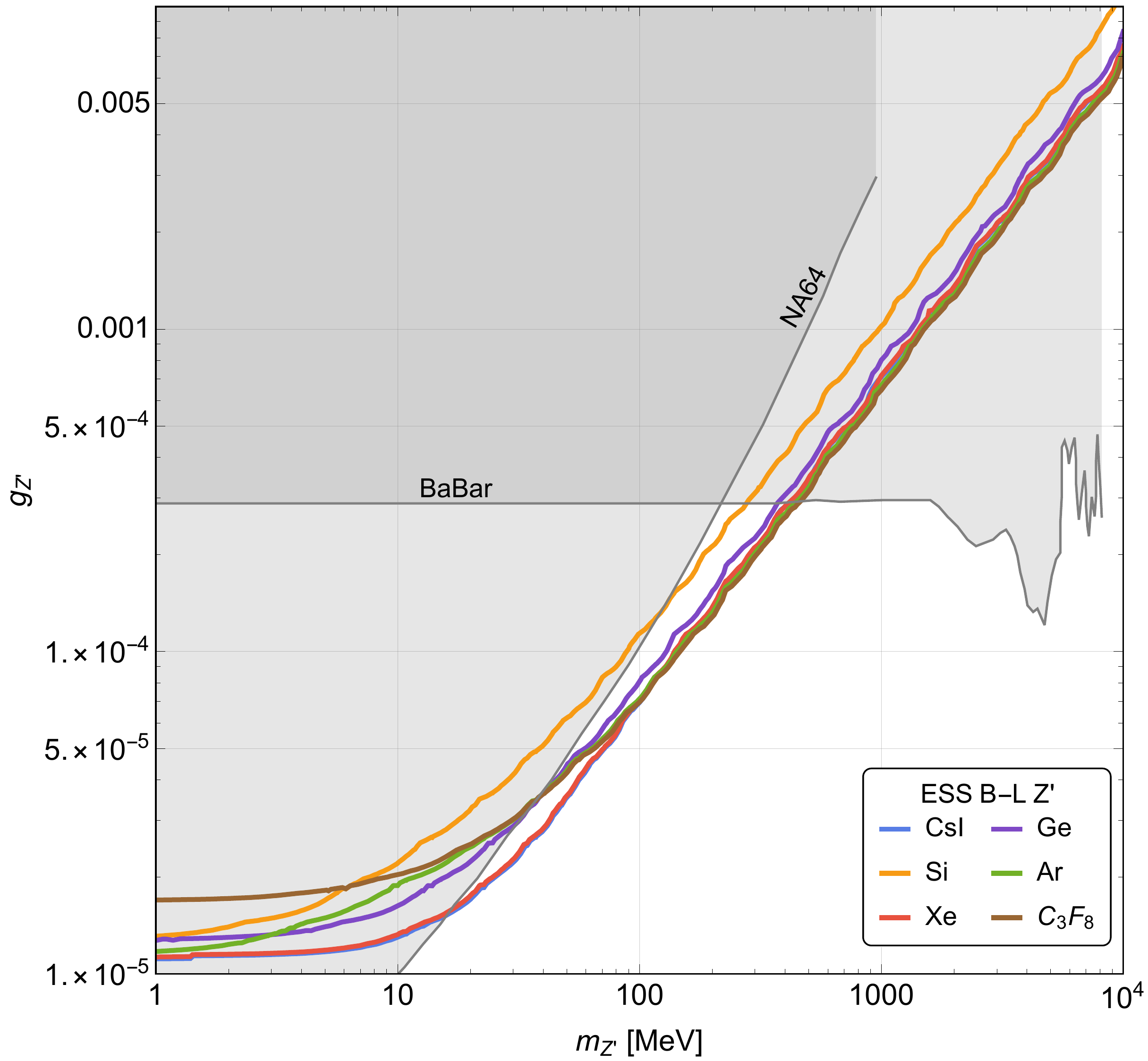}
    \caption{Future sensitivity at $90\%$ C.L. in the $m_{Z'} - g_{Z'}$ plane for the $B-L$ model for the $\nu$BDX-DRIFT detector (left) exploiting CS$_2$ at two different pressures, and the various detectors described in the text at the ESS (right). The gray shaded areas are excluded at $90\%$ C.L. by BaBar \cite{BaBar:2017tiz} and NA64 \cite{Banerjee:2019pds,NA64:2017vtt} assuming that the $Z'$ decays dominantly in invisible dark sector states.\label{fig:B-LZp_dark}}
  \end{center}
\end{figure}

On the other hand, we show in Figure \ref{fig:B-LZp_dark} how the future $\nu$BDX-DRIFT and ESS sensitivity compare to current bounds assuming the $Z'$ to dominantly decay into invisible particles belonging to a dark sector. The gray region is currently bounded by BaBar \cite{BaBar:2017tiz} and NA64 \cite{Banerjee:2019pds,NA64:2017vtt} searches.\footnote{{The $B-L$ model can also be probed in neutrino-electron scattering at DUNE~\cite{Chakraborty:2021apc} and the same is true also for the $L_\mu - L_\tau$ model to be discussed in next section~\cite{Ballett:2019xoj}. The projected sensitivity shows that this search can put bounds of the same order or even outperform the bounds obtained using CE$\nu$NS. In order not to clutter our plots too much, we decided not to show explicitly DUNE's sensitivity.
}} Both sets of detectors have the potential to improve the limits in the range $100 < m_{Z'}/\mathrm{MeV} < 500$ and above $8$ GeV, where the BaBar experiment abruptly loses sensitivity and the bounds from LEP \cite{Fox:2011fx} are too weak.

\subsection{Sensitivity for the $L_\mu-L_\tau$ model}

\begin{figure}[t!]
  \begin{center}
    \includegraphics[width=0.49\textwidth]{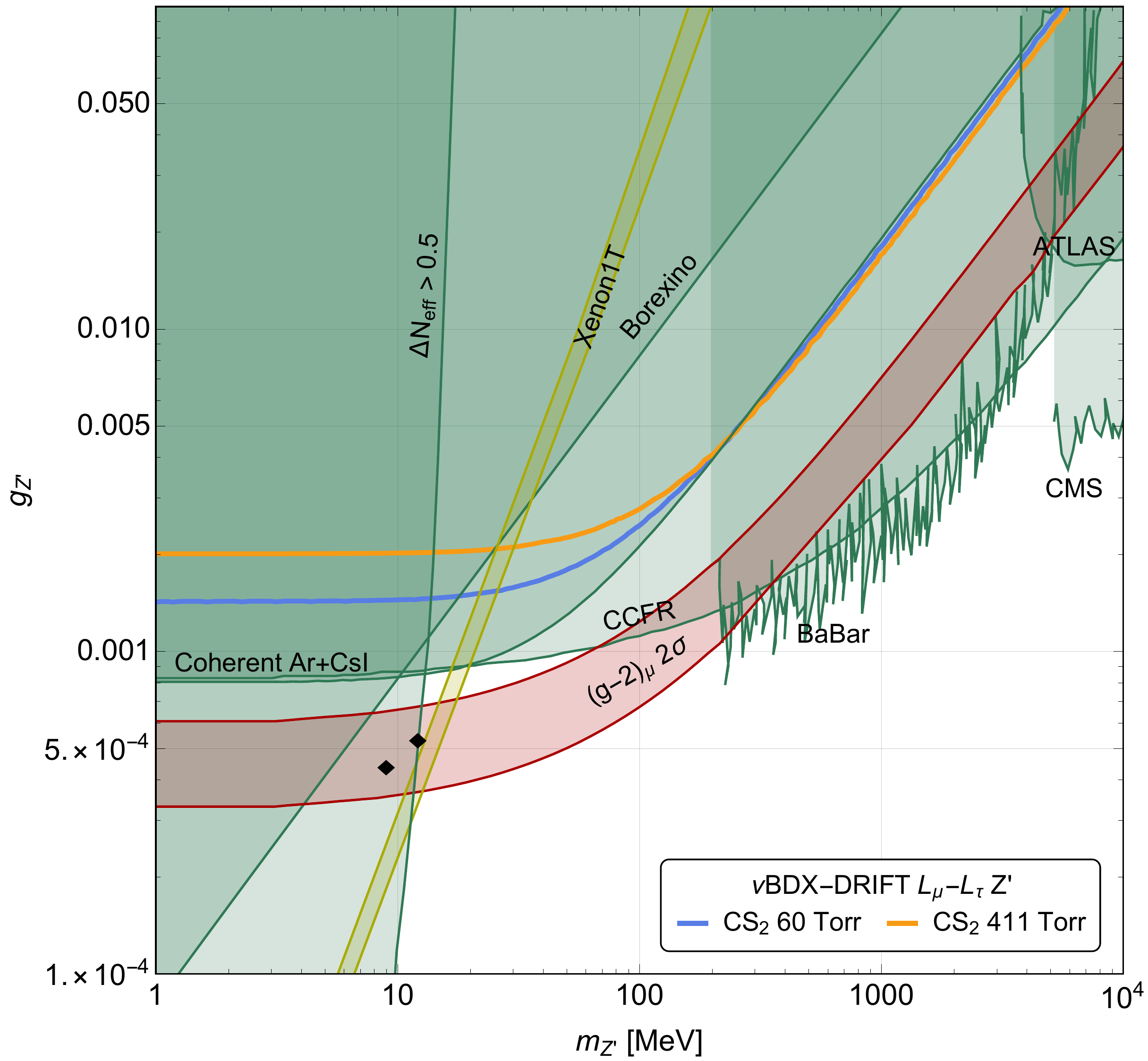}
    \includegraphics[width=0.49\textwidth]{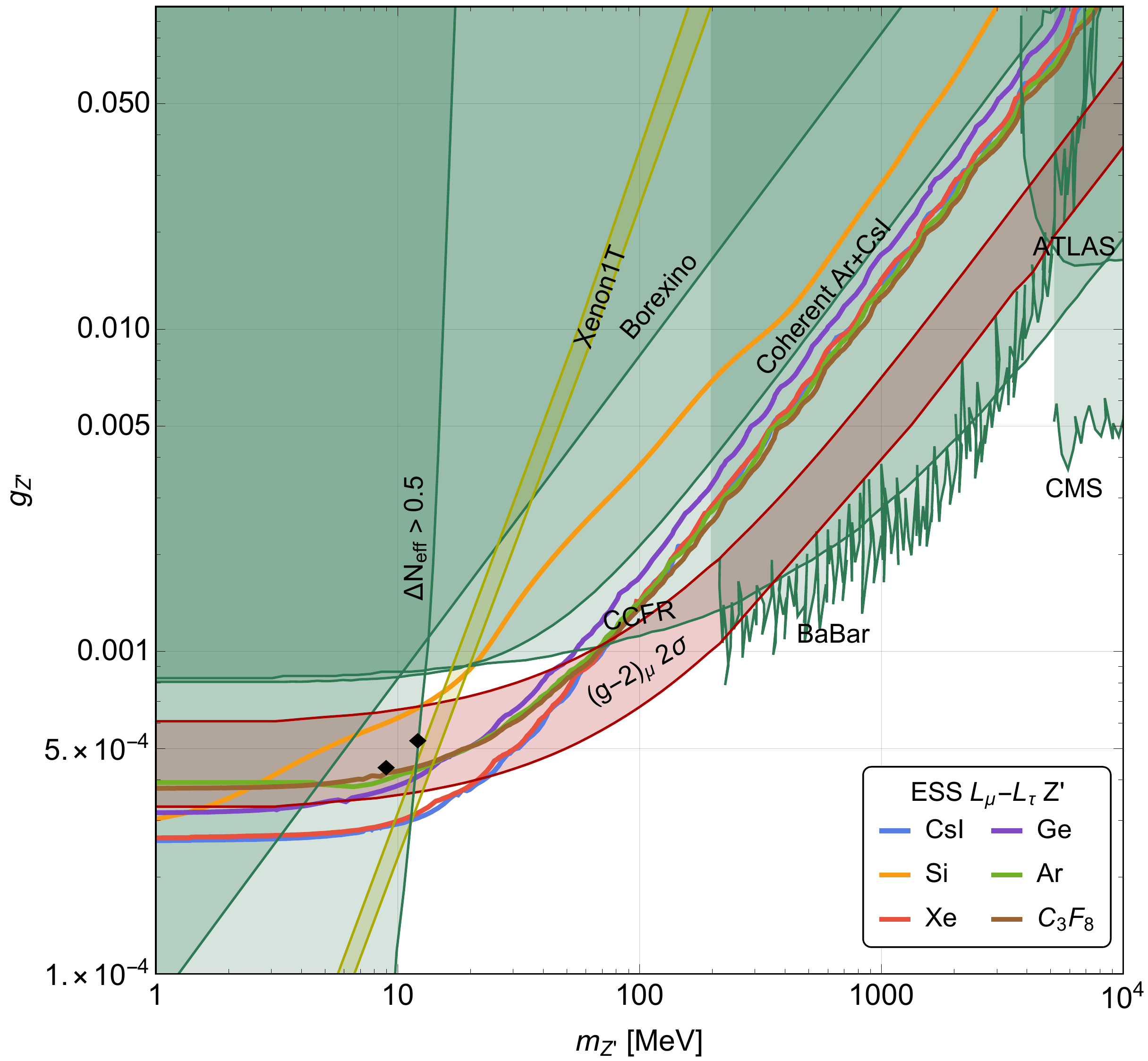}
    \caption{Future sensitivity at 90$\%$ C.L. in the $m_{Z'} - g_{Z'}$ plane for the $L_{\mu}-L_{\tau}$ model for the $\nu$BDX-DRIFT detector (left) exploiting CS$_2$ at two different pressures, and the various detectors described in the text at the ESS (right). The dark green shaded areas are excluded at $95\%$ C.L. by BaBar \cite{BaBar:2016sci}, ATLAS \cite{Altmannshofer:2016jzy,ATLAS:2014jlg}, CMS \cite{CMS:2018yxg}, CCFR \cite{Altmannshofer:2014pba,PhysRevLett.66.3117}, at $90\%$ C.L. by Borexino \cite{Bellini:2011rx,Gninenko:2020xys} assuming that the $Z'$ decays to muons, {and disfavoured by $\Delta N_{\mathrm{eff}}$ \cite{Escudero:2019gzq}}. In the red region the model explains at $2\sigma$ the anomalous magnetic moment of the muon \cite{Muong-2:2006rrc,Aoyama:2020ynm,Muong-2:2021ojo,Davier:2010nc,Davier:2017zfy,Davier:2019can}. The yellow band shows the region that can explain the Xenon1T excess \cite{XENON:2020rca} in some specifically extended $L_\mu-L_\tau$ model \cite{Borah:2021jzu}, while the black diamonds refer to a model \cite{Araki:2015mya} that explain the cosmic neutrino spectrum features observed by IceCube \cite{IceCube:2015gsk}.  \label{fig:LmuLtauZp}}
  \end{center}
\end{figure}

Figure \ref{fig:LmuLtauZp} shows the $90\%$ C.L. sensitivity obtained by the $\nu$BDX-DRIFT detector at the LBNF (left) and several hypothetical experiments at the ESS (right) for the $L_\mu-L_\tau$ model. The sensitivities are compared with the $95\%$ C.L. excluded regions (dark green) obtained in \cite{Altmannshofer:2014pba} exploiting the neutrino trident cross section measured by the CCFR collaboration \cite{PhysRevLett.66.3117}, by the SM $Z$ bosons decaying into four leptons searches at the ATLAS \cite{Altmannshofer:2016jzy,ATLAS:2014jlg} and CMS \cite{CMS:2018yxg} experiments (which can be reinterpreted assuming that the SM Z boson decays into a $Z'$ and two muons), and by the BaBar search for $e^+e^-\to Z\mu^+\mu^-$, where the $Z'$ decays into muons \cite{BaBar:2016sci}. Finally, we show also the reinterpretation by \cite{Gninenko:2020xys} of the Borexino limits \cite{Bellini:2011rx} (dark green) and the $2\sigma$ region needed to explain the anomalous magnetic moment of the muon (red region) \cite{Muong-2:2006rrc,Aoyama:2020ynm,Muong-2:2021ojo,Davier:2010nc,Davier:2017zfy,Davier:2019can}. {The dark green band for $m_{Z'}\lesssim 10$ MeV is disfavoured by the value of $\Delta N_{\mathrm{eff}}$ \cite{Escudero:2019gzq}.} While the $\nu$BDX-DRIFT detector will not be sensitive to regions not already excluded by current existing searches, the various experiments at the ESS have the potential to probe a large part of unexplored region and even to exclude part of the region needed to explain the $(g-2)_\mu$ anomalous magnetic moment of the muon. In particular, these detectors will reach unexplored regions in the range $3< m_{Z'}/\mathrm{MeV}<70$ and may be able to completely exclude the solution to the anomalous magnetic moment of the muon for $m_{Z'}<30$ MeV. Furthermore, it is interesting to note that in specifically extended models \cite{Borah:2021jzu} there is a region which could explain the excess of low energy electrons observed by the Xenon1T experiment \cite{XENON:2020rca} (yellow band) and simultaneously the $(g-2)_\mu$ anomaly (red region). The detectors at the ESS have the potential to probe the values of $(m_{Z'}, g_{Z'})$ that explain both the $(g-2)_\mu$ and the Xenon1T excess, and also the reference points of a specific model \cite{Araki:2015mya} explaining peculiar features observed in the cosmic neutrino spectrum by the IceCube collaboration \cite{IceCube:2015gsk} (also shown in Fig.~\ref{fig:LmuLtauZp}). 

\section{Conclusions}\label{sec:conclusions}

In this paper we discussed the sensitivity that proposed \cevns experiments can reach on light $Z'$ models. More specifically, we have analyzed the $\nu$BDX-DRIFT proposal and studied several detectors that could be installed at the ESS. The following three models have been studied in detail: (i) a universal $Z'$ model in which the light spin-1 particle couples to all the SM fermions with universal strength; (ii) a model in which the $Z'$ couples to the anomaly free $B-L$ current; (iii) a model in which the $Z'$ couples to the anomaly free $L_\mu - L_\tau$ current. Our main results are presented in Figs.~\ref{fig:UniversalZp} -- \ref{fig:LmuLtauZp}, in which we show, together with the sensitivities, existing limits derived from searches at \cevns, fixed target, accelerator, solar neutrino and reactor experiments.

A generic conclusion that can be derived from our study is that the sensitivity of the $\nu$BDX-DRIFT detector is weaker than the one of the experiments at the ESS facility and will test only a small portion of unexplored parameter space. It is however possible to tune the pressure of the gas in order to increase the sensitivity and this may prove important in order to probe blind spot regions. On the other hand, the proposed detectors at the ESS will explore larger portions of untested parameter space. This is particularly true for the $L_\mu-L_\tau$ model where the detector exploiting Xe and CsI will be able to test the region $3\lesssim m_{Z'}/\mathrm{MeV}\lesssim 60$, where the $Z'$ is able to explain the anomalous muon magnetic moment measurements~\cite{Muong-2:2006rrc,Aoyama:2020ynm,Muong-2:2021ojo,Davier:2010nc,Davier:2017zfy,Davier:2019can} and the Xenon1T excess~\cite{XENON:2020rca,Borah:2021jzu}. Moreover, all the detectors at the ESS will be able to test a specific model~\cite{Araki:2015mya} that can explain peculiar features observed in the cosmic neutrino spectrum by IceCube~\cite{IceCube:2015gsk}. 

\section*{Acknowledgements}
E.B. acknowledges financial support from ``Funda\c{c}\~ao de Amparo \`a Pesquisa do Estado de S\~ao Paulo'' (FAPESP) under contract 2019/04837-9.
G.G.d.C. is supported by the INFN Iniziativa Specifica Theoretical Astroparticle Physics (TAsP) and by the Frascati National Laboratories (LNF) through a Cabibbo Fellowship call 2019.
L.M.D.R. acknowledges financial support from ``Conselho Nacional de Pesquisa'' (CNPq) under contract 131297/2020-1.

\bibliography{biblio}
\bibliographystyle{utphys}

\end{document}